\documentclass[twocolumn,prl,showpacs,amsmath,amssymb]{revtex4}

\usepackage{graphicx}
\usepackage{dcolumn}
\usepackage{bm,color}

\begin{document}

\preprint{APS/}

\title{Examination for the stability of a
 rod-shaped structure in $^{24}$Mg}

\author{Yoritaka Iwata$^{1}$}
 \email{iwata@cns.s.u-tokyo.ac.jp}
\author{Takatoshi Ichikawa$^{2}$}
\author{Naoyuki Itagaki$^{2}$}
\author{Joachim A. Maruhn$^{3}$}
\author{Takaharu Otsuka$^{1,4,5}$}

 \affiliation{$^{1}$Center for Nuclear Study, The University of Tokyo, 113-0033 Tokyo, Japan \\
$^{2}$Yukawa Institute for Theoretical Physics, Kyoto University, 606-8502 Kyoto, Japan \\
$^{3}$Institut f\"ur Theoretische Physik, Universit\"at Frankfurt, D-60438 Frankfurt am Main, Germany\\
$^{4}$Department of Physics, The University of Tokyo, 113-0033 Tokyo, Japan. \\
$^{5}$National Superconducting Cyclotron Laboratory, Michigan State University, East Lansing, MI 48824, USA.
}

\date{\today}

\begin{abstract}
The stable existence of rod-shaped structure in highly excited states of $^{24}$Mg is studied based on a systematic cranked Hartree-Fock calculation
with various Skyrme-type interactions.
Its stability is examined by allowing the transition of the cluster like structure to the shell-model like structure.
Especially, the rod-shaped state is exposed to two major instabilities: the bending motion, which is the main path for the transition 
to low-lying states, and the spin-orbit interaction,
which is the driving force to break the $\alpha$ clusters and enhance the independent motion of the nucleons.
The rod-shaped structure with large angular momentum is obtained as a meta-stable stationary state.
\end{abstract}

\maketitle

Strongly deformed nuclear states
with an aspect ratio 1:2 called superdeformed states~\cite{Nyako} have been
found in various nuclei.  The
hyperdeformed states, in which the deformation is around 1:3, have been
also reported in several experiments~\cite{Galindo-Uribarri}.  At first,
those strongly deformed states were found in heavy nuclei,
and whether more exotic states exist in light nuclei under the influence of
$\alpha$-cluster structure is an intriguing question. 

One of the possible candidate is a rod-shaped state in $^{24}$Mg. 
Already at an early stage of the heavy ion physics in the beginning of 1960s, the resonance states of 
$^{12}$C+$^{12}$C around the Coulomb barrier have been discovered~\cite{MR}, 
and their interpretation as the molecular resonance states has been 
extensively investigated. This study has been extended to higher energy regions, and 
since the famous Hoyle state of $^{12}$C at $E_x = 7.65$~MeV has been known as three-$\alpha$ state, 
the six-$\alpha$ cluster states in $^{24}$Mg have been searched. 
Wuosmaa {\em et\ al.}~\cite{Wuosmaa} and Rae {\em et\ al.}~\cite{Rea} suggested that the
molecular resonance of $^{12}$C+$^{12}$C that decays into $^{12}$C($0^+_2$)+$^{12}$C($0^+_2$) 
observed at $E_x = 32. 5$~MeV above the $^{12}$C+$^{12}$C threshold (46.4 MeV above the ground state of $^{24}$Mg) 
was the candidate for such six-$\alpha$ state with rod shape. However it has been shown~\cite{Hirabayashi} 
that this state can be explained within the conventional understanding
of dilute three-$\alpha$ states for each second $0^+$ state of $^{12}$C without introducing
specific geometrical shapes.
No clear evidence has been established despite many efforts.

For the emergence of exotic configuration such as the rod shape,
specific mechanism to stabilize the state is needed. 
Very promising candidate is the rotation of the system;
a large moment of inertia is favored due to the centrifugal force,
when large angular momentum is given to the systems.
Along this line,
Flocard {\em et\ al.} have found that the linear configurations of four-$\alpha$'s appear in large angular momentum states of $^{16}$O~\cite{Flocard}. 
Note here that, while $^{24}$Mg was also studied in Ref.~\cite{Flocard}, the rod-shaped structure of  $^{24}$Mg could not be actually treated due to their restricted model space, and only structures such as $^{12}$C+ $^{12}$C and $^{16}$O+$\alpha$+$\alpha$ were investigated instead.
Although this is an important pioneering work, two major mechanisms for the instability were absent.
One is the bending motion, which is an essential path for the transition to
low-lying states~\cite{Itagaki-C,Maruhn-C,Umar}, and
the other is the spin-orbit interaction, which breaks the $\alpha$ clusters
and enhances the independence of each nucleon~\cite{CS1,CS2}.

It has been found in Ref.~\cite{Ichikawa}
that the four-$\alpha$ rod-shaped configuration can be stabilized
in spite of including these mechanism for the instability. 
The success of four-$\alpha$ rod-shaped structure leads us to the study for
the long-standing issue of nuclear structure physics: six-$\alpha$ rod-shaped state.
However, the presence of stable rod-shaped states is not trivial and
the significance is a cut above the four-$\alpha$ case.
Since the ground state of $^8$Be is considered to have $\alpha$-$\alpha$ cluster configuration,
the four-$\alpha$ rod-shaped state is nothing but an alignment of deformed ground states of $^8$Be. 
On the other hand, the ground state of $^{12}$C is not three-$\alpha$ state~\cite{uegaki,fukuoka}. Even if 
we excite $^{12}$C to the second $0^+$ state with three-$\alpha$ configuration,  
the state is gas-like and does not have any specific shape. 
There have been many efforts to stabilize three-$\alpha$ rod-shaped configurations by adding
valence neutrons~\cite{Itagaki-C,Maruhn-C}; however clear evidence has not been found until now at least 
in experiments, let alone for the six $\alpha$ case.
In this paper, based on the mean field theory, we perform systematic calculations for the rod-shaped configuration in $^{24}$Mg
and show the region of the stability.

\begin{figure} [b]
 \includegraphics[width=8.40cm]{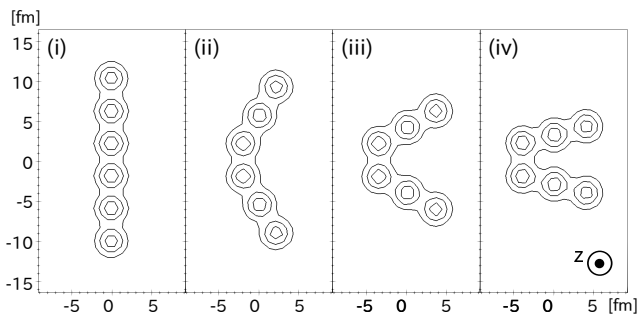}
\caption{Density profiles of the initial states of six-$\alpha$ clusters  
for the cranked Hartree-Fock iteration. 
Panel (i) shows the rod-shaped structure, while panels (ii), (iii), and (iv) correspond to bent structures with opening 
angles of $2\pi/3$, $\pi/3$, and $\pi/6$, respectively. 
Contours are incremented by 0.03~fm$^{-3}$, respectively. 
} 
\label{fig1}
\end{figure}

\begin{figure*} 
 \hspace{-80mm}  (A) \hspace{80mm}  (B)  \\
 \includegraphics[width=7.50cm]{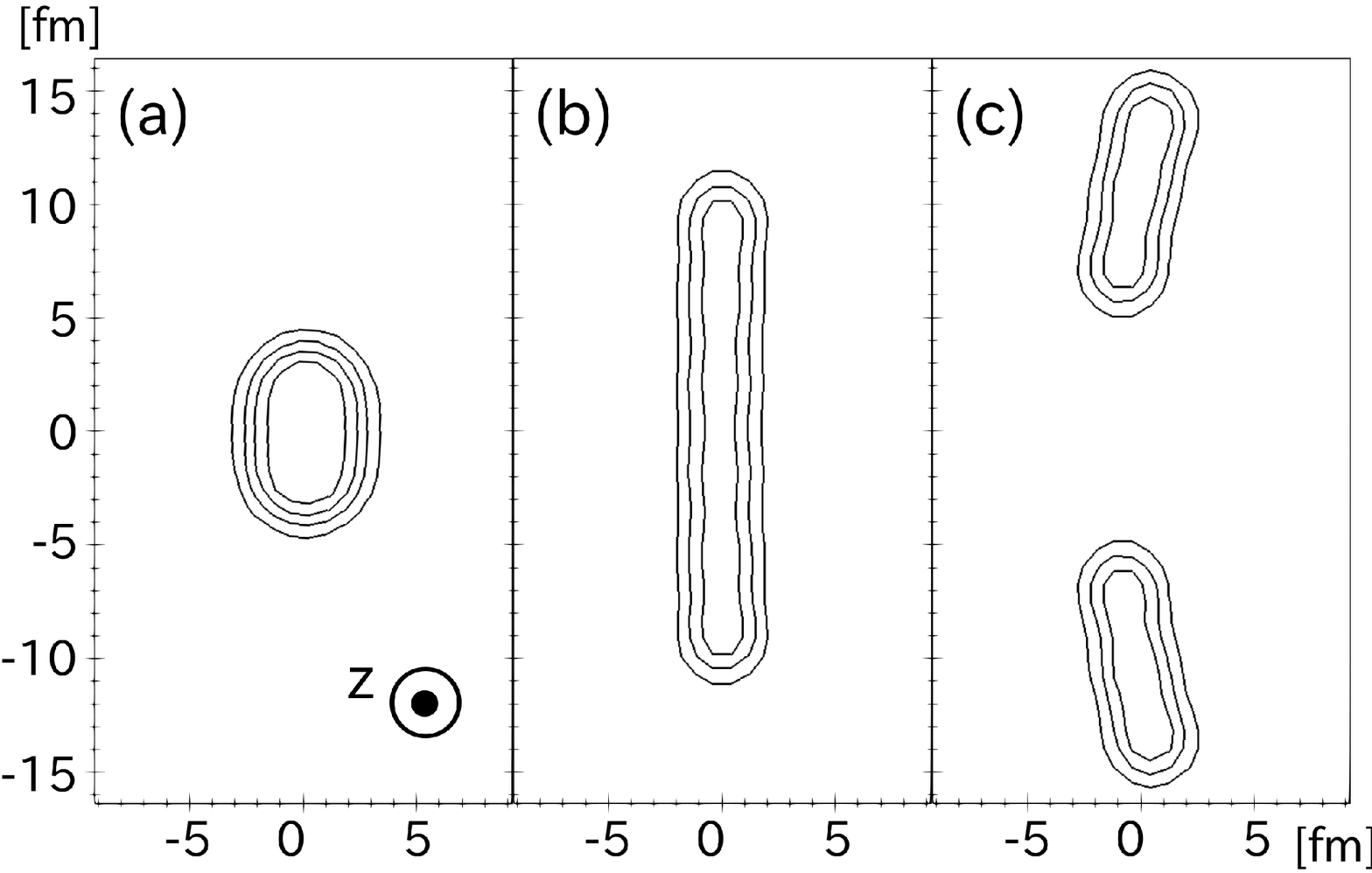} \hspace{20mm}
 \includegraphics[width=7.0cm]{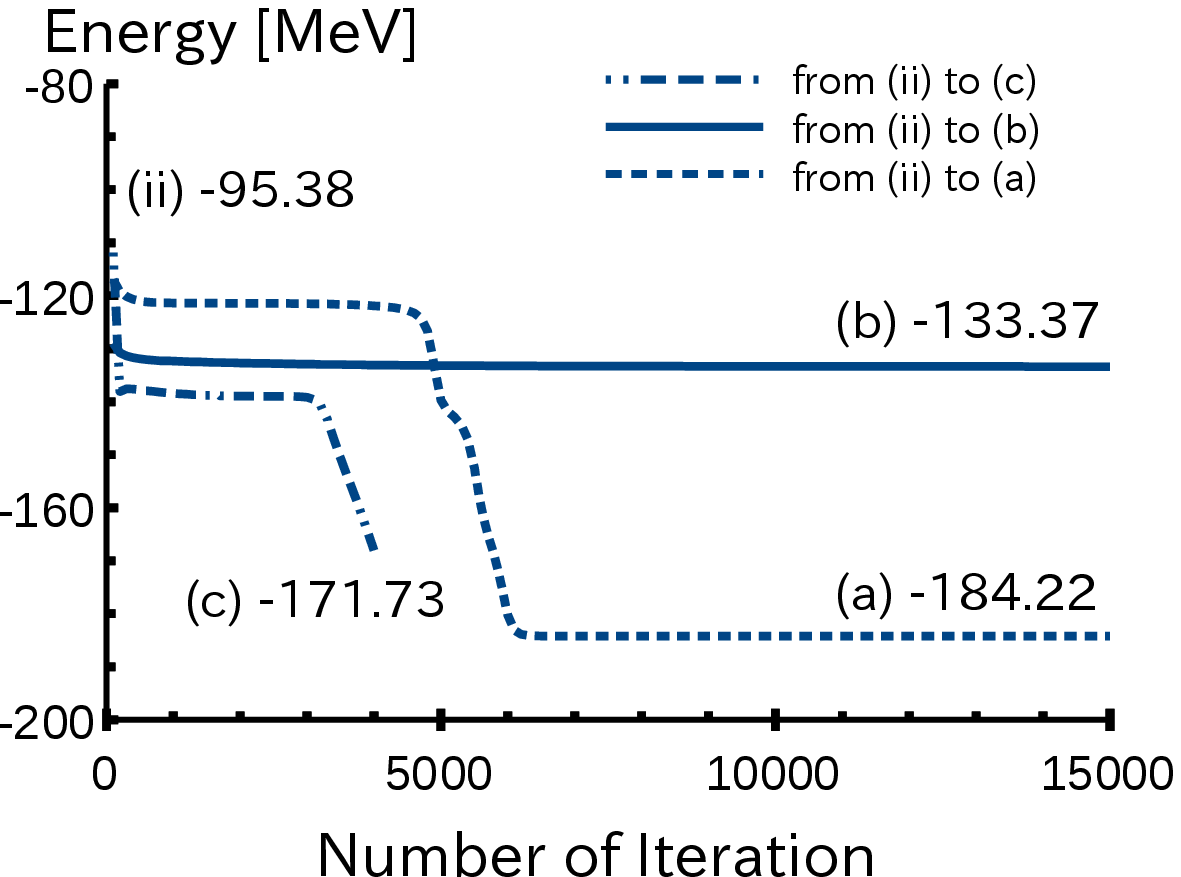}
\caption{(color online) (A) Final states obtained by cranked Hartree-Fock 
calculations (SV-bas parameter set),
(a): compact shape, (b): rod shape, and (c): fragmentation.
Contours are incremented the same as in Fig~\ref{fig1}. 
(B) The imaginary time evolution starting with (ii) of Fig.~\ref{fig1} and $\omega = 0.0$~(MeV/$\hbar$)
going to the final state (a), 
that with (ii) and  $\omega = 1.0$ going to (b), and that with (ii) and  
$\omega = 1.5$ going to (c) (dashed, solid, and dotted-dashed lines).
The expectation values of $H'$ 
(Hamiltonian in the rotating frame)
in Eq.~\eqref{eq1} are shown to see the convergence.} 
\label{fig2}
\end{figure*}

We solve the cranked Hartree-Fock equation in a rotating frame obtained from
the variational principle;
\begin{equation} \label{eq1}  
\delta H' = 0,\ \ \ H'=H - \omega J
\end{equation}
where $H$ is the original Hamiltonian (in the laboratory frame), 
$H'$ is the Hamiltonian in the rotating frame (center of mass frame),
$\omega$ is the rotational frequency, and $J$ is the angular momentum about the $z$-axis.
By solving the equation, the expectation value of the Hamiltonian in the rotating frame is optimized~\cite{Ichikawa}.

Several different initial states for the Hartree-Fock iteration are prepared and
their density distributions are shown in Fig.~\ref{fig1}.
Not only the pure rod-shaped configuration (i), but also other initial states
are prepared by decreasing the opening angle
(opening angles for (ii), (iii) and (iv)
are $2\pi/3$, $\pi/3$ and $\pi/6$ respectively), and we investigate 
how small the opening angle of the initial state can be 
to obtain the rod-shaped configuration after the Hartree-Fock iteration.
Here the $\alpha$ particles are aligned on the $x-y$ plane, and
the rotation axis is chosen to be parallel to the $z$-axis.

\begin{table} 
 \caption{The final states obtained after cranked Hartree-Fock interactions
for different initial states, Skyrme interactions and frequency $\omega$ (MeV/$\hbar$). 
The compact shape, rod shape, and fragmentation are shown by $\circ$, $\bigstar$, and $\ddagger$
corresponding to (a), (b), (c) of Fig.~\ref{fig2} (A), respectively.
\vspace{1.5mm}}
{initial state (i)}
  \begin{tabular}{|c||cccccccccccc|} \hline
   $\omega$   &&  0.0-0.5  & 0.6 & 0.7 & 0.8 & 0.9 & 1.0 & 1.1 & 1.2 & 1.3 & 1.4 & 1.5  \\ \hline 
   SV-bas  &&  $\bigstar$ & $\bigstar$ & $\bigstar$ & $\bigstar$ & $\bigstar$ & $\bigstar$ & $\bigstar$ & $\bigstar$ & $\bigstar$ & $\ddagger$ & $\ddagger$ \\ \hline 
   SV-min  &&  $\bigstar$ & $\bigstar$ & $\bigstar$ & $\bigstar$ & $\bigstar$ & $\bigstar$ & $\bigstar$ & $\bigstar$ & $\bigstar$ & $\ddagger$ & $\ddagger$ \\ \hline 
   SkI3    &&  $\bigstar$ & $\bigstar$ & $\bigstar$ & $\bigstar$ & $\bigstar$ & $\bigstar$ & $\bigstar$ & $\bigstar$ & $\bigstar$ & $\bigstar$ & $\ddagger$ \\ \hline 
   SkI4    &&  $\bigstar$ & $\bigstar$ & $\bigstar$ & $\bigstar$ & $\bigstar$ & $\bigstar$ & $\bigstar$ & $\bigstar$ & $\bigstar$ & $\ddagger$ & $\ddagger$ \\ \hline 
   SLy4    &&  $\bigstar$ & $\bigstar$ & $\bigstar$ & $\bigstar$ & $\bigstar$ & $\bigstar$ & $\bigstar$ & $\bigstar$ & $\bigstar$ & $\ddagger$ & $\ddagger$ \\ \hline 
   SLy6    &&  $\bigstar$ & $\bigstar$ & $\bigstar$ & $\bigstar$ & $\bigstar$ & $\bigstar$ & $\bigstar$ & $\bigstar$ & $\bigstar$ & $\ddagger$ & $\ddagger$ \\ \hline 
   SkM*    &&  $\bigstar$ & $\bigstar$ & $\bigstar$ & $\bigstar$ & $\bigstar$ & $\bigstar$ & $\bigstar$ & $\bigstar$ & $\bigstar$ & $\ddagger$ & $\ddagger$ \\ \hline 
   SkP     &&  $\bigstar$ & $\bigstar$ & $\bigstar$ & $\bigstar$ & $\bigstar$ & $\bigstar$ & $\bigstar$ & $\bigstar$ & $\ddagger$ & $\ddagger$ & $\ddagger$ \\ \hline 
   SkT6    &&  $\bigstar$ & $\bigstar$ & $\bigstar$ & $\bigstar$ & $\bigstar$ & $\bigstar$ & $\bigstar$ & $\bigstar$ & $\bigstar$ & $\ddagger$ & $\ddagger$ \\ \hline 
  \end{tabular}  \vspace{1.5mm} \\ 
 {initial state (ii)}
  \begin{tabular}{|c||cccccccccccc|} \hline
   $\omega$   && 0.0-0.5 & 0.6 & 0.7 & 0.8 & 0.9 & 1.0 & 1.1 & 1.2 & 1.3 & 1.4 & 1.5  \\   \hline 
   SV-bas  &&   $\circ$ & $\circ$ & $\circ$ & $\bigstar$ & $\bigstar$ & $\bigstar$ & $\bigstar$ & $\bigstar$ & $\bigstar$ & $\ddagger$ & $\ddagger$ \\ \hline 
   SV-min  &&   $\circ$ & $\circ$ & $\circ$ & $\bigstar$ & $\bigstar$ & $\bigstar$ & $\bigstar$ & $\bigstar$ & $\bigstar$ & $\ddagger$ & $\ddagger$ \\ \hline 
   SkI3    &&   $\circ$ & $\bigstar$ & $\bigstar$ & $\bigstar$ & $\bigstar$ & $\bigstar$ & $\bigstar$ & $\bigstar$ & $\bigstar$ & $\bigstar$ & $\ddagger$ \\ \hline 
   SkI4    &&   $\circ$ & $\circ$ & $\bigstar$ & $\bigstar$ & $\bigstar$ & $\bigstar$ & $\bigstar$ & $\bigstar$ & $\bigstar$ & $\bigstar$ & $\ddagger$ \\ \hline 
   SLy4    &&   $\circ$ & $\circ$ & $\bigstar$ & $\bigstar$ & $\bigstar$ & $\bigstar$ & $\bigstar$ & $\bigstar$ & $\bigstar$ & $\ddagger$ & $\ddagger$ \\ \hline 
   SLy6    &&   $\circ$ & $\circ$ & $\bigstar$ & $\bigstar$ & $\bigstar$ & $\bigstar$ & $\bigstar$ & $\bigstar$ & $\bigstar$ & $\ddagger$ & $\ddagger$ \\ \hline 
   SkM*    &&   $\circ$ & $\circ$ & $\circ$ & $\circ$ & $\bigstar$ & $\bigstar$ & $\bigstar$ & $\bigstar$ & $\bigstar$ & $\ddagger$ & $\ddagger$ \\ \hline 
   SkP     &&   $\circ$ & $\circ$ & $\circ$ & $\bigstar$ & $\bigstar$ & $\bigstar$ & $\bigstar$ & $\bigstar$ & $\ddagger$ & $\ddagger$ & $\ddagger$ \\ \hline 
   SkT6    &&   $\circ$ & $\circ$ & $\circ$ & $\circ$ & $\circ$ & $\bigstar$ & $\bigstar$ & $\bigstar$ & $\bigstar$ & $\ddagger$ & $\ddagger$ \\ \hline 
  \end{tabular}  \vspace{1.5mm} \\
 {initial state (iii)}
  \begin{tabular}{|c||cccccccccccc|} \hline
   $\omega$   && 0.0-0.5 & 0.6 & 0.7 & 0.8 & 0.9 & 1.0 & 1.1 & 1.2 & 1.3 & 1.4 & 1.5  \\   \hline 
   SV-bas  &&   $\circ$ & $\circ$ & $\circ$ & $\circ$ & $\circ$ & $\circ$ & $\circ$ & $\circ$ & $\bigstar$ & $\ddagger$ & $\ddagger$ \\ \hline 
   SV-min  &&   $\circ$ & $\circ$ & $\circ$ & $\circ$ & $\circ$ & $\circ$ & $\circ$ & $\circ$ & $\bigstar$ & $\ddagger$ & $\ddagger$ \\ \hline
   SkI3    &&   $\circ$ & $\circ$ & $\circ$ & $\circ$ & $\circ$ & $\bigstar$ & $\bigstar$ & $\bigstar$ & $\bigstar$ & $\bigstar$ & $\ddagger$ \\ \hline
   SkI4    &&   $\circ$ & $\circ$ & $\circ$ & $\circ$ & $\circ$ & $\bigstar$ & $\bigstar$ & $\bigstar$ & $\bigstar$ & $\bigstar$ & $\ddagger$ \\ \hline
   SLy4    &&   $\circ$ & $\circ$ & $\circ$ & $\circ$ & $\circ$ & $\circ$ & $\circ$ & $\bigstar$ & $\bigstar$ & $\ddagger$ & $\ddagger$ \\ \hline
   SLy6    &&   $\circ$ & $\circ$ & $\circ$ & $\circ$ & $\circ$ & $\circ$ & $\bigstar$ & $\bigstar$ & $\bigstar$ & $\ddagger$ & $\ddagger$ \\ \hline
   SkM*    &&   $\circ$ & $\circ$ & $\circ$ & $\circ$ & $\circ$ & $\circ$ & $\circ$ & $\circ$ & $\circ$ & $\ddagger$ & $\ddagger$ \\ \hline
   SkP     &&   $\circ$ & $\circ$ & $\circ$ & $\circ$ & $\circ$ & $\circ$ & $\circ$ & $\circ$ & $\ddagger$ & $\ddagger$ & $\ddagger$ \\ \hline
   SkT6    &&   $\circ$ & $\circ$ & $\circ$ & $\circ$ & $\circ$ & $\circ$ & $\circ$ & $\circ$ & $\circ$ & $\ddagger$ & $\ddagger$ \\ \hline
  \end{tabular}    

\label{table1}
\end{table}

The obtained states after the Hartree-Fock iteration depend on
the initial configuration ((i) $\sim$ (iv) of Fig.~\ref{fig1}) and the parameter $\omega$.  
Three typical cases are shown in Fig.~\ref{fig2} (A);
``(a)" (compact shape), ``(b)" (rod shape), and ``(c)" (fragmentation). 
In (b), structure of $\alpha$ clusters are dissolved and necks between $\alpha$'s are formed.
As shown in Fig.~\ref{fig2} (B), if we start with the initial wave function (ii) and 
give $\omega = 0.0, 1.0,$ and $1.50$~MeV/$\hbar$,
we obtain the final states (a), (b), and (c), respectively. 
The iterations toward the final states (a) and (b) converge already in the 10000-th step.
For the solid and dashed lines,
the energy of the rod-shaped states are around $-130$ MeV. 
The jump from this energy to around $-180$ MeV shown in the dashed line corresponds to the change of the shape
from the rod one to the compact one during the Hartree-Fock iteration.
For the dotted-dashed line going to the final state (c), the energy has already been converged at 4000-th step. 
Beyond 4,000-th step numerical error occurs because 
there is no stable solution, and we stop the calculation here.
The energies in this figure are for $H'$, the Hamiltonian in the rotating frame.

In Table~\ref{table1}, the systematics of the obtained final states 
for various Skyrme interactions and $\omega$ values
are summarized.
As a function of $\omega$, we notice 
three kinds of changes of the final state configuration: 
(a)$\leftrightarrow$ (b), (a)$\leftrightarrow$(c), and (b)$\leftrightarrow$(c).
If $\omega = 0$~MeV/$\hbar$ and the initial state is set to (i) (pure rod-shaped state), the final state is (b), and
the rod shape is conserved during the Hartree-Fock iteration independent of the parameter set of the Skyrme interaction;
the rod-shaped structure (even without any rotation) could be a meta-stable stationary state.
By giving finite $\omega$ values, the final states are obtained to be also (b); however we obtain (c), showing fission
at very large $\omega$ values ($\omega > 2.0$~MeV/$\hbar$).
 Here we notice the role of the rotation in the shape transition; an elongated shape is formed due to the centrifugal force,
and fission occurs when the rotation frequency exceeds a certain value.
The compact shape is not obtained starting with the initial configuration (i).

The lowest excitation energy for the rod shape (employing SV-bas) corresponds to $E_x = 78.9$ MeV (initial state (i), $\omega$=0 MeV/$\hbar$).
Note that it is quite reasonable to find the rod shape in much higher energy 
compared with the threshold energy of 6$\alpha$ system ($E_x = 28.5$ MeV in experiments). 
This is because gaslike $\alpha$ cluster states are known to appears 
around the corresponding threshold energies in 3 and 4$\alpha$ systems; however the kinetic energy increases due to uncertainty principle 
if we confine them in one dimensional space. Also, nucleons are excited to very higher nodal orbits
due to the Pauli principle if the degree of freedom is only one dimensional. In addition,
kinetic energy coming from the rotation of the system is quite large.

\begin{figure} 
 \includegraphics[width=5.0cm]{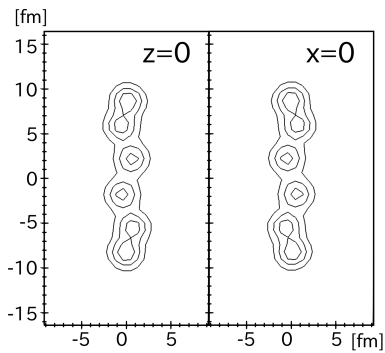}
\caption{An initial state with three-dimensional fluctuation.} 
\label{fig3}
\end{figure}

With decreasing the opening angle of the initial state (changing the initial state as (i)$\to$(ii)$\to$(iii)$\to$(iv)),
the final states gradually change from  the rod shape (b) to the compact shape (a).
Eventually the rod shape is not obtained starting with the initial state (iv).
However, it is worthwhile to mention that the initial state (ii) with the opening angle of $2\pi/3$
gives the rod shape (opening angle $\pi$) in the final state between $\omega$ = 0.9 MeV/$\hbar$ and 1.3 MeV/$\hbar$ independent of the Skyrme interaction parameter sets.
Furthermore, even if we start with the initial configuration (iii),
which has the opening angle of only $\pi/3$,
the final states are still pure rod shapes in the case of some Skyrme interactions.
The results suggest the stability of rod-shaped structure that arises from very strong restoration force against the bending motion depending on the given angular momentum.

\begin{table}
 \caption{Final states starting with the initial state shown in Fig.~\ref{fig3}
for the SV-bas parameter set. 
 Final states with compact shape, rod shape, and fragmentation are shown by (a), (b), and (c), respectively.
 Here $\omega$ is the rotation frequency (MeV/$\hbar$) and values in the parentheses
are the energies (MeV), which are equal to the expectation values of $H$ 
(Hamiltonian in the laboratory frame) in Eq.~\eqref{eq1}.
\vspace{1.5mm}}
  \begin{tabular}{|c||c|c|c|c|c|c|c|c|c|c|c|c|c|c|c|c|c|c|c} \hline
   $\omega$  & 1.0 & 1.1& 1.2 & 1.3 & 1.4 \\   \hline 
  & (a) ($-182.2$) & (b) ($-104.9$)  & (b) ($-100.5$) & (b) ($-94.5$)  & (c)   \\   \hline
  \end{tabular}  \vspace{1.5mm} \\
 \label{table2}
\end{table}

Although all the initial states up to this point were planar configurations,
here we examine the initial six-$\alpha$ configuration with a
three-dimensional distortion as shown in Fig.~\ref{fig3}.
Indeed, as shown in Table~\ref{table2}, the final state with rod shapes is obtained for $\omega$ = 1.1, 1.2, 1.3 and 1.4~MeV/$\hbar$.
It was also found for the other Skyrme interaction parameter sets shown in
Table~\ref{table1} that there exists
a stable region of rod-shaped structure around $\omega = 1.2$~MeV/$\hbar$,
even if this three-dimensional distortion is taken into account in the initial state.

\begin{table*}
 \caption{A comparison of energies (MeV) calculated with (With LS) and without (Without LS) the spin-orbit interaction for given rotational frequency $\omega$(MeV/$\hbar$).
The energies are the expectation values of $H$ in Eq.~\eqref{eq1}
(Hamiltonian in the laboratory frame).
The initial state is (ii) of Fig.~\ref{fig1}, and the SV-bas parameter set is used.
Final states are either compact shapes 
(-185 to -183 MeV)
or rod shapes 
(-114 to -94 MeV).
\vspace{1.5mm}}
  \begin{tabular}{|c||c|c|c|c|c|c|c|c|c|c|c|c|c|c|c|c|c|c|c|c|c} \hline
   $\omega$ &0.0 & 0 .1 & 0.2 & 0.3 & 0.4& 0.5 & 0.6 & 0.7 & 0.8 & 0.9 & 1.0 & 1.1 & 1.2 & 1.3   \\   \hline 
 With LS & $-184.2$  & $-184.2$  & $-184.1$ & $-184.0$  & $-183.9$  & $-183.7$ & $-183.5$  & $-183.3$   & $-113.6$ & $-111.2$ & $-108.3$ & $-104.9$  & $-100.5$  & $-94.5$  \\   \hline
Without LS & $-117.2$ & $-117.0$ & $-116.7$  & $-116.1$ & $-115.3$  & $-114.1$ & $-112.7$ & $-110.9$  & $-108.8$ & $-106.2$ & $-103.1$ & $-99.3$  & $-94.2$  & $-86.2$    \\   \hline
  \end{tabular}  \vspace{1.5mm} \\
 \label{table3}
\end{table*}

\begin{figure} [t]
 \includegraphics[width=8.0cm]{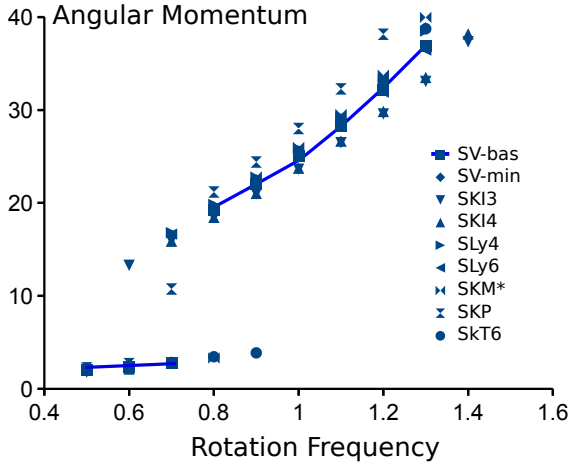}
\caption{(color online) The angular momentum $J$ as a function of rotation frequency $\omega$ (MeV/$\hbar$) for 9 Skyrme interaction parameter sets, where the initial state is set to  (ii) of Fig.~\ref{fig1}.
The line with solid squares denote the calculated result for SV-bas parameter set.} 
\label{fig4}
\end{figure}
The angular momentum of the system as a function of $\omega$ is summarized in Fig.~\ref{fig4},
where the results for the final states (a) and (b) of Fig.~\ref{fig2} are plotted.
In all Skyrme interaction parameter sets the angular momentum jumps;
no significant deformation and resulting angular momentum in the small $\omega$ region, 
and strongly deformed rod shapes with large angular momenta in the large $\omega$ region.
In the case of SV-bas parameter set,
there appears a jump between $\omega = 0.7$ and 0.8~MeV/$\hbar$ 
corresponding to the shape transition of the final state from (a) to (b), 
and the rod shapes give large angular momenta, around $J = 19 \sim 37 \hbar$. 
The points corresponding 
to the rod shape slightly deviate from the straight line, indicating further increase of deformation at large $\omega$ values due to the centrifugal force.

Next, we discuss the effect of spin-orbit interaction.
Since we consider rotations with high angular momentum,  the
effect should be  large.
The spin-orbit interaction  was ``on" in the calculation up to this point,
and here we study its effect by switching it off and  
comparing the results. 
The initial state is set to (ii) of Fig.~\ref{fig1}, and 
the final state is obtained to be either 
``(a)" (compact shape) or ``(b)" (rod shape)
of Fig.~\ref{fig2}
for $\omega$ in the range of 0.0 to 1.3~MeV/$\hbar$.
As shown in Table~\ref{table3},
the spin-orbit interaction strengthens the binding, and the total energies get higher without it.
We notice a clear transition of the final state energy from around $-183$ MeV to $-113$ MeV,
corresponding to the change of the obtained final state from (a) to (b). 
Without the spin-orbit interaction, the region of the rod shapes is
increased to $\omega = 0.0 \sim 1.3$ MeV/$\hbar$, thus showing the
enhanced stability of the $\alpha$ clusters;
however, even with the spin-orbit interaction, the six-$\alpha$-type rod shape has a region of existence as we have shown previously.

When the rod shape is formed, the contribution of the spin-orbit interaction 
(the difference between with and without) is rather small. 
This could be due to the
formation of $\alpha$ clusters; however another interpretation is possible.
The state has a hyperdeformed structure and the nuclear density is very low. The spin-orbit interaction has a derivative of nucleon density and small spin-orbit contribution could be due to this low-density effect.

\begin{figure} 
 \includegraphics[width=8.0cm]{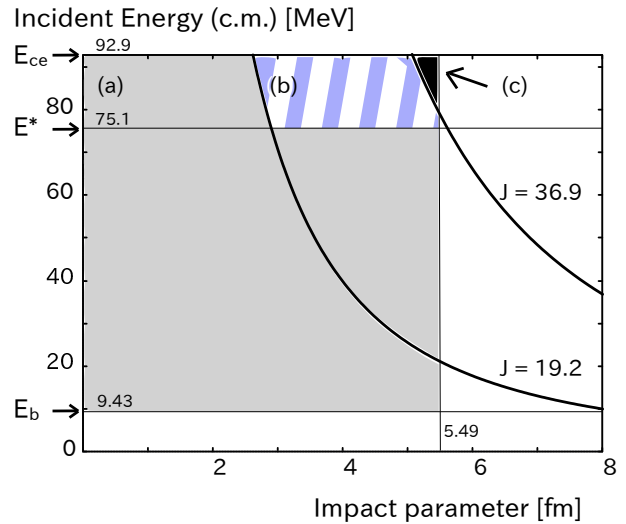}
\caption{(color online) The energy-impact parameter diagram for the appearance of different shapes based on Fig.\ref{fig4}.
The regions for the appearance of different shapes ((a), (b), and (c) of Fig.~\ref{fig2}) are translated 
into the impact parameter and center of mass energy of the $^{12}$C+$^{12}$C reaction (SV-bas parameter set). 
The perpendicular axis means the energy measured from $^{12}$C+$^{12}$C at infinite distance.
$E_{\rm b} = 9.43$~MeV denotes the fusion barrier, and the rod-shaped structure is obtained in the region of and $E = 75.1-94.16$~MeV.
$E^* = 75.1$~MeV at $\omega$ = 0.8~MeV/$\hbar$ is the smallest excitation energy of the rod shape obtained in the structure calculation.
$E_{\rm ce}$ is the charge equilibrium upper limit energy~\cite{iwata}, above which emission of charge non-equilibrated particle may occur.
}
\label{fig5}
\end{figure}

Here we discuss the region for the rod shape 
on an energy-impact parameter diagram.
Figure 5 maps the calculated excitation energies and angular momenta
of rod shaped structure to the corresponding incident energies and
impact parameters of $^{12}$C+$^{12}$C collision. The cranking results tells us
that the incident energy must be larger than 75.1 MeV (the lowest excitation energy at $\omega = 0.8 MeV/\hbar$ for the rod shape
measured from the $^{12}$C+$^{12}$C threshold) and the angular
momentum must be in between 19.2 and 36.9$\hbar$ for the formation of
the rod shape.
Note that the energy 75.1 MeV corresponds to 86.7 MeV if it is measured from the ground state of $^{24}$Mg.
The impact parameter must be smaller than 5.49 (twice
the $^{12}$C radius) so that the $^{12}$C+$^{12}$C reactions occurs. 
As a result, rod shape potentially formed in the striped region (b) in Fig. 5.
Note that this is necessary condition for the formation of the rod shape and not the sufficient condition.

In summary, the existence of rod-shaped structure in highly excited states of $^{24}$Mg 
has been studied based on a systematic cranked Hartree-Fock calculation.
Its stability against both bending motion and the $\alpha$-particle dissociation effect arising mostly from the spin-orbit 
force has been confirmed regardless of the choice of the Skyrme interaction parameter set.
Such rod shape appears in the region of $E_x = 86.7$ MeV, and
it is quite reasonable to find them in higher energy compared with the 6$\alpha$ threshold energy;
although gaslike $\alpha$ cluster states are known to appears 
around the corresponding threshold energies in 3 and 4$\alpha$ systems, the kinetic energy increases
due to uncertain principle 
if we confine them them in one dimensional space. Also, nucleons are excited to very higher nodal orbits
due to the Pauli principle if the degree of freedom is only one dimensional.
The rod shape is quite stable when large angular momentum is given to the system.
Such rotating rod-shaped states, which are considered to be a meta-stable stationary state, are predicted and could be produced in heavy-ion collisions. 

This work was supported in part by MEXT SPIRE and JICFuS, by the Helmholtz alliance HA216/EMMI, and by BMBF under contract No.\ 05P12RFFTG.
Numerical computation was carried out at the Yukawa Institute Computer Facility.


\begin{thebibliography}{50}
 \bibitem{Nyako} 
	 B.M. Nyak\'{o} {\it et al.}, Phys. Rev. Lett. {\bf 52}, 507 (1984).
	 
 \bibitem{Galindo-Uribarri}
	 A. Galindo-Uribarri {\it et al.}, Phys. Rev. Lett. {\bf 71}, 231 (1993).
 	 
\bibitem{MR}
E. Almqvist, D. A. Bromley, and J. A. Kuehner,
     Phys. Rev. Lett. {\bf 4}, 515 (1960).

 \bibitem{Wuosmaa}
	 A.H. Wuosmaa {\it et al.},
	 Phys. Rev. Lett. {\bf 68}, 1295 (1992).
	 
 \bibitem{Rea}
	 W.D.M. Rae, A.C. Merchant, and B. Buck, Phys. Rev. Lett. {\bf 69}, 3709 (1992).
	 
	 
 \bibitem{Hirabayashi}
	 Y. Hirabayashi, Y. Sakuragi, and Y. Abe, Phys. Rev. Lett. {\bf 74},  4141 (1995).

	 
\bibitem{Flocard} 
 H. Flocard, P.H. Heenen, S.J. Krieger, and M. S. Weiss,
 Prog. Theor. Phys. {\bf 72}, 1000 (1984). 

\bibitem{uegaki}
E. Uegaki, S. Okabe, Y. Abe, and H. Tanaka,
Prog. Theor. Phys., {\bf 07}, 4 (1977) 

 \bibitem{fukuoka}
	 Y. Fukuoka, S. Shinohara, Y. Funaki, T. Nakatsukasa, and K. Yabana, Phys. Rev. {\bf C 88}, 014321 (2013).

\bibitem{Itagaki-C}
	N. Itagaki, S. Okabe, K. Ikeda, and I. Tanihata, Phys.  Rev. C {\bf 64}, 014301 (2001);
	 N. Itagaki, W. von Oertzen, and S. Okabe, 
	 {\it ibid}, {\bf 74},  067304 (2006).
 \bibitem{Maruhn-C}
	 J.A. Maruhn, N. Loebl, N. Itagaki, and M. Kimura,
	 Nucl. Phys. A {\bf 833}, 1  (2010).

\bibitem{Umar}
	 A.S. Umar, J.A. Maruhn, N. Itagaki, and V.E. Oberacker,
	 Phys. Rev. Lett. {\bf 104},  212503 (2010).
	 
	 \bibitem{CS1}
 N. Itagaki, S. Aoyama, S. Okabe, and K. Ikeda,
 Phys. Rev. C {\bf 70}, 054307 (2004).
	 
	 \bibitem{CS2}
	 N. Itagaki, H. Masui, M. Ito, and S. Aoyama,
 Phys. Rev. C {\bf 71}, 064307 (2005);
	 Tadahiro Suhara, Naoyuki Itagaki, J\'{o}zef Cseh, and Marek P{\l}oszajczak,
 {\it ibid}, {\bf 87}, 054334 (2013). 
 
 \bibitem{Ichikawa}
	 T. Ichikawa, J. A. Maruhn, N. Itagaki, and S. Ohkubo, 
 Phys. Rev. Lett. {\bf 107}, 112501 (2011).

 \bibitem{iwata}
	 Y. Iwata, T. Otsuka, J. A. Maruhn, and N. Itagaki,
 Phys. Rev. Lett. {\bf 104}, 252501 (2010).

\end{thebibliography}
\end{document}